**Poacher-population dynamics when legal trade of naturally deceased organisms funds anti-poaching enforcement**

Matthew H. Holden[1*], Jakeb Lockyer[1]

1. The University of Queensland, School of Mathematics and Physics,

    St Lucia, QLD, 4072, Australia

*corresponding author: m.holden1@uq.edu.au

**Abstract:**

Can a regulated, legal market for wildlife products protect species threatened by poaching? It is one of the most controversial ideas in biodiversity conservation. Perhaps the most convincing reason for legalizing wildlife trade is that trade revenue could fund the protection and conservation of poached species. In this paper, we examine the possible poacher-population dynamic consequences of legal trade funding conservation. The model consists of a manager scavenging carcasses for wildlife product, who then sells the products, and directs a portion of the revenue towards funding anti-poaching law enforcement. Through a global analysis of the model, we derive the critical proportion of product the manager must scavenge, and the critical proportion of trade revenue the manager must allocate towards increased enforcement, in order for legal trade to lead to abundant long-term wildlife populations. We illustrate how the model could inform management with parameter values derived from the African elephant literature, under a hypothetical scenario where a manager scavenges elephant carcasses to sell ivory. We find that there is a large region of parameter space where populations go extinct under legal trade, unless a significant portion of trade revenue is directed towards protecting populations from poaching. The model is general and therefore can be used as a starting point for exploring the consequences of funding many conservation programs using wildlife trade revenue.

**Introduction**

Illegal harvest of plants and animals (also known as poaching) is one of the greatest threats to biodiversity (Maxwell, S. L. et al. 2016). One of the most commonly proposed solutions to reduce poaching is to ban wildlife trade and punish those who break the law (IUCN 2000; Sadovy de Mitcheson et al. 2018). Yet, trade bans have failed to stem the poaching crisis for several species (Conrad 2012). Some scientists, governments and managers have argued that the solution is to lift bans, and introduce a highly regulated, legal trade of wildlife products. Theoretically, legal trade could increase poached species population sizes by (1) flooding an, otherwise illegal, market with legal product, driving down prices, and therefore decreasing the financial incentive to poach (Walker and Stiles 2010; Biggs et al. 2013) and (2) by generating sales revenue to fund the conservation of poached species (Stiles 2004; Di Minin et al. 2015; Smith et al. 2015). It is one of the most controversial theories in conservation biology (Biggs et al. 2017; Sekar et al. 2018; Eikelboom et al. 2020), eliciting strong supporting (Cooney and Jepson 2006) and opposing arguments (Litchfield 2013; Lusseau and Lee 2016; Sekar et al. 2018). While mathematical models have been used to demonstrate why flooding the market with legal products may benefit (Gentry et al. 2019) or harm (Crookes and Blignaut 2015) different types of poached species, the potential use of legal trade revenue to fund conservation has largely been ignored by mathematical biologists, despite the unique nonlinear feedbacks this policy would create.

These nonlinear feedbacks can be explained by thinking about how wildlife products gaining in value affect poaching. When products command high prices, poachers make more money, increasing their incentive to harvest more organisms. But if the government is also selling the product and using the revenue to fund conservation, the government is also able to sell the product for higher prices as well. This generates extra conservation funding, which may benefit the species. Alternatively, poachers have less financial incentive to poach when prices are low, but low prices also mean smaller conservation budgets. Without the aid of mathematical models, this trade-off could be spun negatively or positively for those who argue for legal trade. In one sense, if conservation funding went towards anti-poaching law enforcement, the opposing forces could help prevent increased poaching when prices increase. But, in many models, forces that appear to be stabilising can lead to periodic orbits, where populations cycle between high and low sizes, risking extinction during bad years, due to random effects. This is why it is important to rigorously analyze explicit mathematical models of legal trade funding conservation.

In this paper we propose a simple system of ordinary differential equations that captures the essence of legal trade funding biological conservation. Because, in many instances, it is morally and politically infeasible for a government to harvest live charismatic animals (Biggs et al. 2017), we consider a mathematical model where the manager can only sell wildlife product derived from animals who die naturally. If governments could find animal carcasses fast enough, this is potentially feasible for many non-perishable wildlife products, such as, teeth, bones, ivory, scales, and horns. Once the legal product is scavenged from carcasses it is then sold alongside the illegal product derived from animals killed by poachers. The price of the product is determined by the supply of the product on the market, and poachers increase their poaching effort when it is profitable, and decrease effort when poaching is unprofitable. The government uses a portion of their sales revenue to increase anti-poaching law enforcement, which increases the expected cost of poaching, decreasing poaching profitability. In this paper we examine the long-term population consequences of this potential policy and derive the critical proportion of legal trade revenue to be allocated towards increasing law-enforcement in order for the population to no longer be threatened by poaching.

**Poacher-population dynamic model**

Let $x(t)$ be the population density or abundance of the harvested species at time $t$. Note, from now on we will frequently omit the "$(t)$" to make the equations cleaner. Population size changes through time according to the model,

$$\frac{dx}{dt} = (b - m)x - qxy, \qquad (1)$$

with intrinsic population growth rate, $b - m$, which is the difference between a birth rate, $b$, and a mortality rate, $m$. It is necessary to track naturally dying individuals through an explicit mortality rate because our goal is to examine the population dynamics of poached species when a manager scavenges deceased organisms to sell wildlife product. The second term on the right denotes that poachers decrease the population size of the poached species by illegally harvesting individual organisms, where $y(t)$ is a measure of poaching effort, at time $t$, and the parameter $q$ is the catchability of organisms by poachers. We assume the rate at which poachers change their effort, is proportional to the profitability of poaching,

$$\frac{dy}{dt} = \alpha\,[\,p(x,y)qxy - c(x,y)y\,], \qquad (2)$$

where $p(x,y)$, is the price paid to poachers per poached individual, and $c(x,y)$ is the cost of poaching per unit poaching effort. Equation (2) is the standard way of modelling harvest effort in theoretical population biology (Hilborn et al. 2006; Mansal et al. 2014; Holden et al. 2018; Crookes and Blignaut 2019). It says poachers increase effort when poaching is profitable and decrease effort when it is unprofitable, with the parameter $\alpha$ controlling how fast poachers adjust effort. We assume price, $p(x,y)$, declines with increased wildlife product on the market, through a demand curve (Marshall 1890), from economics,

$$p(x, y) = \frac{p_0}{1 + a(qxy + smx)}. \qquad (3)$$

The parameter $p_0$ is the maximum price any buyer is willing to pay for the product, and $a$ determines how much the price decreases with increased product on the market (solid black curve in Fig. 1). Market supply consists of product entering the market from poaching, $qxy$, and legal sales, $smx$, where $s$ is the proportion of carcasses the manager is able to scavenge for legal trade. We assume both illegal and legal products command the same price, to simplify the analysis. The functional form of equation (3) is different from most price models in the literature. Price is more commonly assumed to decrease linearly with increasing supply (Clark 1990; van Kooten 2008; Auger et al. 2010; Chen 2016). While this linear curve is appropriate in some cases, it can lead to negative prices (dashed red curve in Fig. 1), because the government may sell product even when it is unprofitable for poachers to poach. While this is unrealistic, we derive analogous results for our model if equation (3) was replaced with a linear price function, in the supplement, with a summary on pgs. 17-18. Another standard approach is modelling price as a function of one over the supply (Ly et al. 2014; Burgess et al. 2017; Fryxell et al. 2017), for which the Hill equation, one over supply raised to a fixed power, is the most common choice, because it guarantees fixed price elasticity of demand (Fryxell et al. 2017). Unfortunately, the specific nonlinear structure of this equation makes a global analysis of the model presented in this paper analytically intractable, and additionally, in such a model, prices grow to infinity as the supply approaches zero (dot-dashed blue curve in Fig. 1). Our alternative price curve (3), which captures many features of the Hill equation, namely strictly positive, convex, decreasing price as a function of supply is therefore an ideal substitute (solid black curve in Fig. 1). While (3) is less commonly used in the literature than the Hill equation, it is often presented in introductory economics textbooks (Sydsaeter et al. 2002).

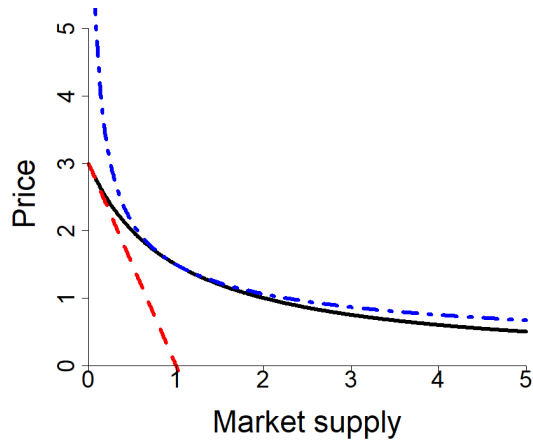

**Fig. 1.** Price as a function of supplied wildlife product on the market for equation 3 (solid black curve), for linear price, according to equation 13, (dashed red curve), and price according to the Hill equation, $(\gamma/[smx + qxy])^\beta$ (dash-dotted blue curve). Parameters are $p_0 = 3$, $a = 1$, $\gamma = 9/4$, $\beta = 1/2$.

The cost per unit of poaching effort is modelled as,

$$c(x, y) = c_O + c_F \lambda(x, y), \qquad (4)$$

where, $c_O$, is the opportunity cost of poaching, [e.g. lost income from not doing other activities]. The second term is the expected cost of getting caught poaching, which is proportional to the density of law enforcement, $\lambda(x,y)$. The parameter $c_F$ can be thought of as the expected rate of money paid by poachers, per unit poaching and law enforcement effort, from being caught poaching. Note the payment rate could be from explicit fines, or nonmonetary costs, such as jail terms, but converted into dollars. We assume law-enforcement effort is

$$\lambda(x,y) = \lambda_0 + gfsmxp(x,y). \tag{5}$$

Where, $\lambda_0$, is some baseline level of law-enforcement the manager deploys from an external budget and where the second term represents the increase in law-enforcement due to legal trade revenue. Because $s$ is the proportion of naturally deceased organisms legally scavenged, $smx$ is the rate of individual organisms legally being supplied to the market. Therefore, $smxp(x,y)$ is the rate of money flowing to the government from legal sales. The parameter $f$ is the proportion of that money the government chooses to allocate to enforcement, and, $g$, converts the money into additional enforcement, via enforcement costs. Note that species body size is implicitly accounted for in parameters, $b$, $m$, and $p_0$, as larger bodied organisms tend to have lower birth and death rates and produce more wildlife product per individual. However, explicit size and age structure of the population is ignored here as a simplifying assumption.

In the previous model, to simplify the analysis, we assumed linear population growth in the absence of harvest. But in reality, populations do not grow to infinity. Therefore, we introduce a model where population growth slows as the population exhausts its resources,

$$\frac{dx}{dt} = \left[\frac{bmk}{mk + (b-m)x} - m\right]x - qxy, \tag{6}$$

where, $k$ is the carrying capacity of the population. Here, per-capita birth rate (the first term in the bracket) is equal to $b$ at zero population density, and then monotonically decreases to zero as population density goes to infinity, according to a hyperbolic function. The parameterization of the hyperbolic decline is chosen such that the birth rate equals the death rate when $x = k$. In other words, it is a model that generates a stable equilibrium population abundance equal to carrying capacity, in the absence of harvest. As is standard in ecology, we include density dependence in the birth rate, since for the vast majority of species, newborns are the most vulnerable life stage (Quinn II and Deriso 1999). Most mathematical biologists will be more familiar with the logistic equation, $rx(1 - x/k)$, as a way to incorporate density dependent population growth. This equation will not work for our purposes, because we must tease apart the mortality rate from the birth rate, in order to track wildlife products created from deceased organisms. While $(b - m)x(1 - x/k)$ might appear reasonable, in such a model, the mortality rate is actually $m + bx/k$. Since we are interested in tracking mortality, equation (6) is preferable, because *m* maintains its interpretation as the natural mortality rate, as in the linear population growth model.

**Assumption A1:** *parameters $b, m, q, \alpha, c_O, c_F, g, p_0, k$ and $a$, are strictly positive real numbers and additionally, $\lambda_0 \geq 0$, $b > m$, and $f, s \in [0,1]$.*

This assumption guarantees that the population grows in the absence of harvest, that poachers can successfully catch some individuals from the population, and that poachers always increase poaching effort when profitable and decrease poaching effort when unprofitable. It also guarantees positive prices for the wildlife product, and some non-zero cost of deploying poaching effort. The case, $a = 0$, corresponds to a fixed price, $p_0$, per unit of wildlife product, meaning price does not change with respect to market supply. In such a scenario, the model reduces to the classic Lotka-Voltera model, where poachers are the predators and the harvested population is the prey. The dynamics of the Lotka-Voltera model, closed periodic orbits (Kot 2003), are well known, and hence we restrict our study to the case where $a > 0$.

**Nondimensionalized Models**

The linear population growth model (equations 1-5) can be non-dimensionalized, and written more simply as

$$\frac{dX}{d\tau} = X - XY, \quad (7)$$

$$\frac{dY}{d\tau} = \left(\frac{uX}{1 + wX + XY} - v\right)Y, \quad (8)$$

which has only three parameters, $u$, $w$, and $v$, and three nondimensionalized state variables $X$, $Y$, and $\tau$, which are all given as follows,

$$X = a(b-m)x, \qquad Y = \frac{qy}{b-m}, \qquad \tau = (b-m)t,$$

$$u = \frac{\alpha p_0(q - c_F fgms)}{a(b-m)^2}, \qquad w = \frac{ms}{b-m}, \qquad v = \frac{\alpha(c_O + c_F \lambda_0)}{b-m}.$$

Note an alternative rescaling of the model would leave a parameter on the $XY$ term in the denominator, rather than the $w$, on the $X$ term. We chose the above rescaling because when $w = 0$, one recovers the model under the case where $s$ and $f$ are zero. A similar non-dimensionalization of the nonlinear population growth model (equations 2 - 6) yields,

$$\frac{dX}{d\tau} = \left(\frac{1-X}{1+nX}\right)X - XY \quad (9)$$

$$\frac{dY}{d\tau} = \left(\frac{\psi X}{1 + \omega X + zXY} - v\right)Y \quad (10)$$

where,

$$X = \frac{x}{k}, \qquad n = \frac{b-m}{m}, \qquad \psi = \frac{\alpha k p_0(q - c_F fgms)}{b-m},$$

$$\omega = amsk, \qquad z = ak(b-m),$$

with $Y$, $\tau$ and $v$ the same as in the linear case.

**Poacher-population dynamics**

In this section, we highlight the possible population and poacher dynamics in each of the models. We demonstrate that if a condition on the parameters is satisfied, the harvested population either grows to infinity, in the linear population growth model, [equations 7 and 8], or to carrying capacity, in the nonlinear model, [equations 9 and 10]. If the condition is not satisfied, poacher and population numbers approach a stable equilibrium. Therefore, the condition on the parameters provides a useful management tool for projecting the effects of using legal trade to fund anti-poaching law-enforcement on wildlife populations. The key results for managers and conservation biologists are equations 11 and 13, which translate the condition on the parameters into a critical proportion of legal trade revenue that must be allocated to anti-poaching law enforcement to achieve abundant wildlife populations.

The advantage of the linear population model is that it leads to mathematically tractable dynamics. In this model, not only can we classify population dynamic behaviour near equilibria, but also classify the global behaviour of trajectories given any initial condition and parameter combination. While a global analysis is less tractable for the nonlinear population growth model in equation (9), we illustrate, numerically, and through local linear stability analysis, that this model produces qualitatively similar dynamics to the linear population growth model in equation (7).

In the linear model, equations (7) and (8), with assumption A1, there are three possible types of qualitative dynamics, as depicted in Fig. 2. If $v < \frac{u}{1+w}$, there is a stable positive equilibrium $(X^*, Y^*) = \left(\frac{v}{u-v[1+w]}, 1\right)$, where population size and poacher effort approach $(X^*, Y^*)$, given any initial condition with some non-zero poaching effort and population size (Fig. 2ab). Otherwise, if $v \geq \frac{u}{1+w}$, the population grows to infinity for any non-zero initial population size (Fig. 2c). The stability condition, $v < \frac{u}{1+w}$, can be rewritten as $u - v[1+w] < 0$. If, $u - v[1+w] \geq \frac{v^3 w^2}{4u}$, then the stable equilibrium, $(X^*, Y^*)$, is a spiral, (see Fig 2a), whereas if, $0 < u - v[1+w] < \frac{v^3 w^2}{4u}$, then $(X^*, Y^*)$ is a stable node, (see Fig 2b). When, $u - v[1+w] = 0$, there is a bifurcation where the denominator of $X^*$ is zero, denoting the loss of the non-trivial equilibrium from the positive quadrant.

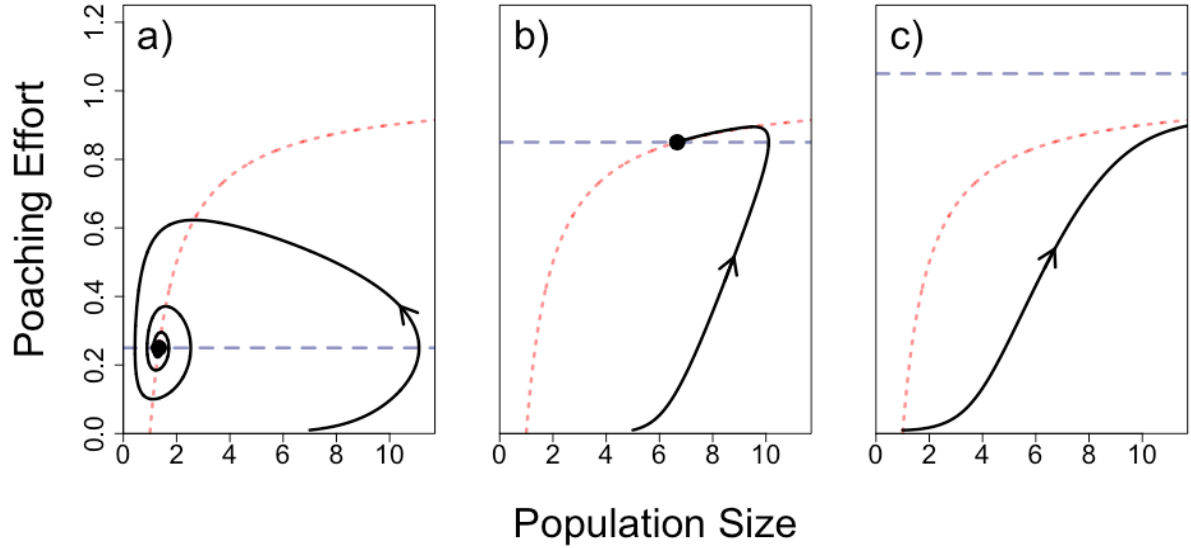

**Fig. 2.** The three types of possible dynamics, in the phase plane, given equations (7) and (8). a) In the case where, $0 < u - v[1+w] < \frac{v^3 w^2}{4u}$, there is a stable spiral attracting all strictly positive initial conditions. b) When $u - v[1+w] > \frac{v^3 w^2}{4u}$, the equilibrium is a stable node. c) When $u - v[1+w] \leq 0$, the nontrivial equilibrium vanishes and population size grows indefinitely, to infinity. The parameters are $w = 0$ for all plots. In a) $u = 2.768, v = 2.353$, in b) $u = 768, v = 2.353$, and in c) $u = 1.814, v = 1.905$. These parameters correspond to $q = p_0 = c_O + c_F \lambda_0 = a = 1$, and $f = s = 0$, in the original dimensional model, equations (1-5), and additionally with, $\alpha = 1$, $b - m = 0.25$ in a) and $\alpha = 2$, $b - m = 0.85$ in b) and $\alpha = 2$, $b - m = 1.05$ in c). The dotted, red curve is the non-trivial, $dy/dt = 0$, nullcline, and the blue dashed curve is the non-trivial, $dx/dt = 0$, nullcline. All values in the plot are of the dimensional system, equations (1–5), for ease of biological interpretation.

The only other equilibrium is a trivial equilibrium (0,0), which is always a saddle. It has a stable manifold corresponding to an initial population size of zero, where poaching effort then decreases to zero because it is always unprofitable to try and poach a non-existent species. The unstable manifold of (0,0) corresponds to zero initial poaching, where the population grows indefinitely in the absence of harvest. The possible dynamics, described above, can be summarized more formally by the following theorem.

**Theorem 1**: *For any initial condition* $(X(0), Y(0)) \in \{(X,Y): X > 0, Y > 0\}$, *and given* $\frac{dX}{d\tau}$ *and* $\frac{dY}{d\tau}$ *according to equations* (6,7), *and assumption A1, if* $v < \frac{u}{1+w}$, *then*

$$\lim_{\tau \to \infty} X(\tau) = \frac{v}{u-v[1+w]} \text{ and } \lim_{\tau \to \infty} Y(\tau) = 1. \text{ If } v \geq \frac{u}{1+w}, \text{ then } \lim_{t \to \infty} X(t) = \infty \text{ and}$$

$$\lim_{t \to \infty} Y(t) = \frac{u}{v} - w.$$

**Proof**: Let $v < \frac{u}{1+w}$, then the non-negative equilibria are (0,0) and $\left(\frac{v}{u-v[1+w]}, 1\right)$. It is straightforward to show, through a standard linear stability analysis, that (0,0) is an unstable saddle with stable manifold $\{(X,Y): X = 0\}$ and unstable manifold $\{(X,Y): Y = 0\}$, and that $\left(\frac{v}{u-v[1+w]}, 1\right)$ is a stable node for $0 < u - v[1+w] < \frac{v^3 w^2}{4u}$, and stable spiral for $u - v[1+w] > \frac{v^3 w^2}{4u}$. See Appendix 1 for detailed linear stability analysis calculations.

Therefore, $\left(\frac{v}{u-v[1+w]}, 1\right)$ is locally stable if $v < \frac{u}{1+w}$. The *Bendixon-Dulac theorem* can be used to show there are no periodic orbits in the positive quadrant, by choosing $B(X,Y) := \frac{1}{XY}$, and noting that $\nabla \langle B \frac{dX}{d\tau}, B \frac{dY}{d\tau} \rangle = -\frac{uX}{(1+wX+XY)^2}$, which is always negative for $X > 0$. Therefore, $\{(X,Y): X > 0, Y > 0\}$ is the basin of attraction for $\left(\frac{v}{u-v[1+w]}, 1\right)$.

If $v \geq \frac{u}{1+w}$, the saddle $(0,0)$ is the only equilibrium satisfying $X \geq 0$ and $Y \geq 0$.

Additionally, $\frac{dX}{d\tau} < 0$ for all $(X(\tau), Y(\tau)) \in \{(X,Y): Y > max\left(0, \frac{u}{v} - w - \frac{1}{X}\right), X > 0\}$. Since $\frac{dX}{d\tau} > 0$, for all $(X(\tau), Y(\tau)) \in \{(X,Y): X > 0, Y < 1\}$ and $\frac{u}{v} - w - \frac{1}{X} < 1$, for all $X > 0$, it follows, from continuity of the vector field, that $\lim_{t \to \infty} X(t) = \infty$ and $\lim_{t \to \infty} Y(t) = \frac{u}{v} - w$.

Theorem 1 allows us to analyze the dynamics in the rescaled, dimensional model, equations 1-5, which are easier to interpret biologically. For example, in the case where there is no legal trade, $f = s = 0$, the dynamics presented above can be written in terms of the population growth rate, $b - m$. Low population growth rates, $b - m < \frac{p_0 q}{ac}\left(\frac{4 p_0 q}{\alpha a c^2 + 4 p_0 q}\right)$, produce the stable spiral (Fig 2a), intermediate population growth rates, $\frac{p_0 q}{ac}\left(\frac{4 p_0 q}{\alpha a c^2 + 4 p_0 q}\right) < b - m < \frac{p_0 q}{ac}$, lead to the stable node (Fig 2b), and large population growth rates, $b - m \geq \frac{p_0 q}{ac}$, cause the nontrivial equilibrium to vanish, with population size growing to infinity (Fig 2c). This makes sense, as we would expect larger long-term population sizes for more quickly growing species.

Theorem 1 also allows us to derive the critical law enforcement funding rate, in order for legal trade of naturally deceased organisms to produce abundant wildlife population sizes in the long-term. This is summarized in the following corollary.

**Corollary 1**: *Define,*

$$f^* := \frac{p_0 q - a(b-m)(c_O + c_F \lambda_0)}{c_F g m p_0 s} - \frac{a(c_O + c_F \lambda_0)}{c_F g p_0}, \qquad (11)$$

*and*

$$x^* := \frac{c_O + c_F \lambda_0}{a(c_O + c_F \lambda_0)(b - m + ms) + p_0(c_F f gms - q)}. \tag{12}$$

*For any initial condition $(x(0), y(0)) \in \{(x, y): x > 0, y \geq 0\}$, and given $\frac{dx}{dt}$ and $\frac{dy}{dt}$ according to equations $(1-5)$, and assumption A1, if $f \geq f^*$ then $\lim_{t \to \infty} x(t) = \infty$, for all $0 < s \leq 1$. If $f < f^*$ then, $\lim_{t \to \infty} x(t) = x^*$.*

**Proof:** Follows from rescaling the condition $v \geq \frac{u}{1+w}$, and the equation for the equilibrium, $\left(\frac{v}{u-v[1+w]}, 1\right)$, in theorem 1.

In the case where $0 \leq f^* < 1$, then $\lim_{t \to \infty} x(t) = \infty$, as long as $f > f^*$. If, $f^* > 1$, then it is impossible, regardless of the choice of the funding proportion, for legal sales to lead to infinite population sizes. If $f^* < 0$, then the population grows to infinity, even if none of the revenue from legal sales funds anti-poaching law enforcement. This means that either flooding the market with legal product drives the product price down low enough for poaching to be unprofitable, or poaching was never profitable enough to harm the population, even in the absence of legal sales. In Fig. 3, we plot the bifurcation diagram in the system with respect to the two parameters the manager controls, *f* and *s*. The black curve denotes the critical *f*, *f\**, as a function of *s*. The white region below the curve is the region of parameter space where the population approaches the stable equilibrium. The light grey region above the curve is the region where introducing some positive value for *f*, above the curve, causes infinite growth. The darker grey region, to the right of the intersection of the curve, with *f* = 0, denotes the region where the population would grow to infinity regardless as to whether legal sales funded enforcement or not. For example, with parameters as in Fig 3, when scavenging 20 percent of dead organisms for wildlife products, $s = 0.2$, increasing the percent of revenue towards law-enforcement funding from zero to 100 percent does not lead to infinite population sizes (circle a, and d, in Fig 3, and plots a and d in Fig 4). However, when $s = 0.5$, moving from $f = 0$, to $f = 1$, leads to infinite populations sizes, as $f$ crosses the critical $f^*$ value (circle b, and e, in Fig 3, and plots b and e in Fig 4). Alternatively, the population always goes to infinity if $s = 0.9$, no matter the value for $f$ (circle c, and f, in Fig 3, and plots c and f in Fig 4).

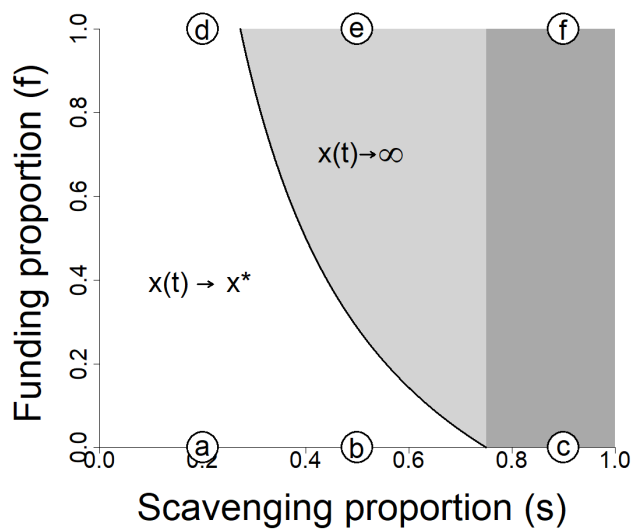

**Fig. 3:** Bifurcation diagram for the ODEs described in equations 1-5. Other parameters are $b = 2, p_0 = 3.5, m = q = c_o = c_F = \lambda_0 = g = a = 1$. In the white region, the population approaches a stable equilibrium. In the light grey region ($f$, as a function of $s$ above the critical black curve, $f^*$) the population size approaches infinity. In the dark grey region, the population grows to infinity, regardless as to the choice of $f$. The phase diagrams for the corresponding circles in this bifurcation diagram are plotted in Figure 3. For example, for $s = 0.5$ and $f = 0$, see figure 3b. At the same scavenging efficiency, $s = 0.5$, but allocating all revenue to buying extra law-enforcement, $f = 1$, see figure 3e, where the extra funding leads to population sizes which grow to infinity.

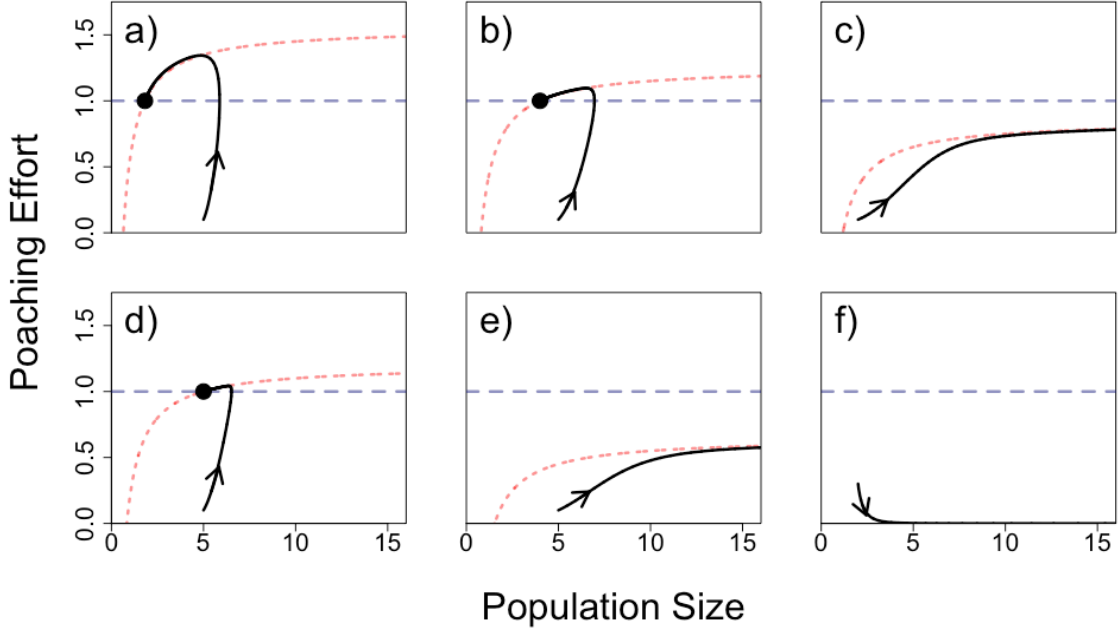

**Fig. 4:** Phase diagrams for the ODEs described in equations 1-5 (in the top row, zero legal sales revenue is allocated to funding enforcement, $f = 0$, bottom row all legal sales revenue is allocated to law enforcement $f = 1$). Columns correspond to different scavenging efficiencies (a,d, $s = 0.2$, b,e, $s = 0.5$, c,f, $s = 0.8$). The labels correspond to the circles in $(s, f)$ parameter space labelled on the bifurcation diagram in Fig. 3. Other parameters are $b = 2, p_0 = 3.5, m = q = c_O = c_F = \lambda_0 = g = a = 1$.

We show in the supplement that replacing equation 3 with a price function that decreases linearly with supply,

$$p(x, y) = p_0[1 - a(qxy + smx)], \tag{13}$$

produces similar dynamics to those shown above, and there is an analogous condition to that presented in corollary 1 for the critical funding proportion which guarantees infinitely growing populations, namely,

$$f^*_{linear} := \frac{p_0 q - 4a(b-m)(c_O + c_F \lambda_0)}{c_F g m p_0 s} - \frac{4a(c_O + c_F \lambda_0)}{c_F g p_0}. \tag{14}$$

The only difference between equation (11) and equation (14) is the multiplication by four in the terms on the right. This means we require a smaller proportion of legal revenue allocated towards enforcement to guarantee infinitely growing populations if we assume price declines linearly with supply. Therefore, equation (11) can be thought of as a more conservative condition.

There is one major difference between the qualitative dynamics when the nonlinear price equation, (3), is replaced by the linear equation, (13), in our model. This is the introduction of a second equilibrium, a saddle, for the case where $f < f^*_{linear}$. The saddle equilibrium occurs at a larger population size than the stable positive equilibrium. So for linear price curves, large initial population sizes grow to infinity, while small initial populations flow towards the smaller equilibrium. This is mostly due to the possibility of negative prices (see the grey area in Fig. S1 in the supplement). If the government scavenges from populations growing indefinitely, negative prices are guaranteed as the market supply approaches infinity. So poachers rapidly decrease poaching to avoid negative revenue. The introduction of negative prices is unrealistic, as the government would surely stop selling scavenged product if prices went negative. Therefore, we prefer the price curve in equation (3) as a more theoretically justified mode for our particular application.

All previous results assumed that the population growth rate was linear with respect to population size in the absence of poaching. The linear population model was especially useful because we could prove the global behaviour of all trajectories, analytically, including ruling out periodic orbits. Adding nonlinearity to population growth makes proofs of the global dynamics less tractable. However, it is possible to explain some of the long-term qualitative behaviour in this model, both analytically and numerically. There are potentially three biologically relevant equilibria in this model. The zero-poacher zero-population size equilibrium still exists and continues to be a saddle. The nullcline, where $dx/dt$ is zero (no population growth), now declines with population size and intersects the $x$ axis at carrying capacity, which is rescaled to be one in the non-dimensionalized model (see blue dashed curve in Fig 5a-c). This intersection corresponds to an equilibrium where there is no poaching and the population is at its maximum sustainable size.

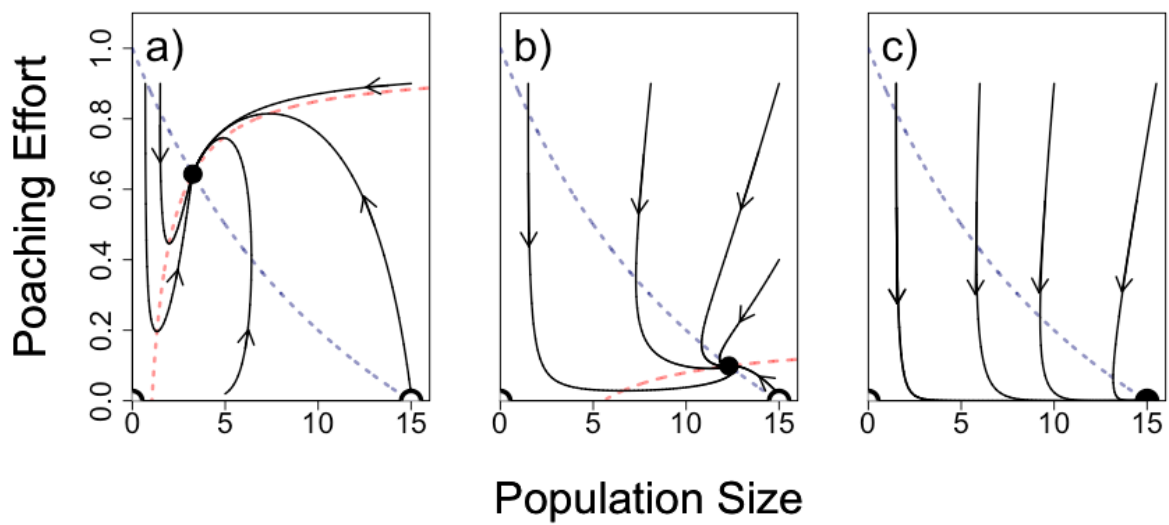

**Figure 5:** Phase diagrams for the ODEs described in equations 2-6. In a) zero legal sales revenue is allocated to funding law-enforcement, $f = 0$, in b) $f = 0.55$, and in c) all legal sales revenue is allocated to law-enforcement, $f = 1$. The labels correspond to the circles in $(s, f)$ parameter space labelled on the bifurcation diagram in figure 6. Other parameters are $m = q = c_O = c_F = \lambda_0 = g = a = 1, b = 2, p_0 = 3.5, \alpha = 3, k = 15, s = 0.8$.

A particular relevant goal from a management perspective, is whether we can eliminate poaching by introducing legal trade. Fortunately, even though the complete global dynamics are difficult to prove mathematically, we can prove a sufficient condition for the existence of this important equilibrium being globally stable in the relevant positive quadrant.

**Theorem 2**: *For any initial condition* $(X(0), Y(0)) \in \{(X, Y): X > 0, Y \geq 0\}$, *and* $\frac{dX}{d\tau}$ *and* $\frac{dY}{d\tau}$ *according to equations* (8,9), *and given assumption A1, if* $v \geq \frac{\psi}{1+\omega}$, *then* $\lim_{\tau \to \infty} X(\tau) = 1$ *and* $\lim_{\tau \to \infty} Y(\tau) = 0$.

**Proof:** Let $v \geq \frac{\psi}{1+\omega}$. Then the only non-negative equilibria are $(0,0)$ and $(1,0)$, because the $\frac{dX}{d\tau} = 0$ nullcline, $Y = \frac{1-X}{1+nX}$, monotonically decreases from $(1,0)$ to $(0,1)$ and the $\frac{dY}{d\tau} = 0$ nullcline, $Y = \frac{u-vw}{vz} - \frac{1}{zX}$, is non-positive for all $X \in (0,1)$. It is straightforward to show, through a standard linear stability analysis, that $(0,0)$ is an unstable saddle with stable manifold $\{(X,Y): X = 0\}$, and unstable manifold $\{(X,Y): Y = 0\}$. The unstable manifold of $(0,0)$ flows directly into $(1,0)$. A linear stability analysis around $(1,0)$ shows it is a stable node because the eigenvalues of the linear system are, $\frac{-1}{1+n}$, and $\frac{u}{1+\omega} - v$ at $(1,0)$. By index theory there can be no periodic orbits in the region $\{(X, Y): X > 0, Y \geq 0\}$. By continuity of the vector field $\{(X, Y): X > 0, Y \geq 0\}$ is the basin of attraction for $(1,0)$.

This allows us to derive a critical *f*, which we will call $\phi$, for the nonlinear model, analogous to equation (11) in the linear model.

**Corollary 2**: *Define,*

$$\phi := \frac{p_0 q - (1/k)(c_O + c_F \lambda_0)}{c_F g m p_0 s} - \frac{a(c_O + c_F \lambda_0)}{c_F g p_0}. \tag{15}$$

*If $f \geq \phi$, and $\frac{dx}{dt}$ and $\frac{dy}{dt}$ are according to equations $(2-6)$, and assumption A1, then For any initial condition $(x(0), y(0)) \in \{(x, y): x > 0, y \geq 0\}$, $\lim_{t \to \infty} x(t) = k$, for all $0 < s \leq 1$. If $f < \phi$, then there exists a $T > 0$, such that for all $t > T$, and $(x(0), y(0)) \in \{(x, y): x > 0, y > 0\}$, $x(t) < k$.*

**Proof:** If $f \geq \phi$, then $\lim_{t \to \infty} x(t) = k$ follows from Theorem 2, after re-dimensionalizing the condition $v \geq \frac{\psi}{1+\omega}$. Alternatively, if $f < \phi$, then $(k, 0)$ is an unstable saddle with its stable manifold not in $\{(x, y): x > 0, y > 0\}$, and because $\frac{dx}{dt} < 0$, for all $x < k$, then $(x(0), y(0)) \in \{(x, y): x < k, y > 0\}$ is invariant, and all trajectories with initial conditions in $(x(0), y(0)) \in \{(x, y): x \geq k, y > 0\}$ flow into the invariant region.

Note that in corollary 1, given the linear model, we proved that the population always approaches the stable non-trivial equilibrium. In corollary 2, we only proved that the population is eventually maintained below carrying capacity. The stability of the nontrivial equilibrium, in the nonlinear model, is difficult to prove analytically, as the formula for the eigenvalues from linear stability analysis around the non-trivial equilibrium involve over one hundred nonlinear terms. However, for all parameter values and initial conditions tested in the positive upper half plane, when $f < \phi$, numerical solutions did indeed approach the non-trivial equilibrium (Fig. 5). In figure 6, we compare the bifurcation diagrams demarcated by the critical funding proportion, $\phi$, and $f^*$, in the nonlinear and linear models respectively. Note that $\phi$, and $f^*$, in equations (11) and (13) are similar. Equation (13) for $\phi$, only differs from $f^*$ by replacing $a(b-m)$ in equation (11) with $1/k$. As a result, these critical curves, that demarcate the different regions in the bifurcation diagram, have the same shape. Large values of $k$ increase the critical proportion of funding towards law enforcement required to achieve stable population size at carrying capacity, $k$. This makes sense because it is easier for poachers to catch the species if it is common. Similarly, it can be shown that as long as it is profitable to poach a population at carrying capacity with no legal sales revenue allocated to enforcement, then $\partial\phi/\partial m, \partial\phi/\partial c_F$ and $\partial\phi/\partial a < 0$. This means that legally selling product from deceased animals is most likely to work for species with high death rates (meaning the population provides more wildlife products to scavenge), governments with high anti-poaching enforcement efficiency (meaning funding extra patrols is effective), and high price sensitivity (meaning flooding the market with product reduces poacher incentives).

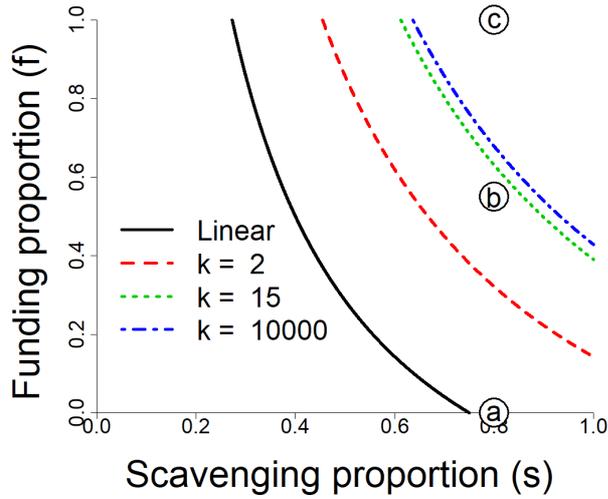

**Fig. 6.** Bifurcation diagram in the linear population growth model and the nonlinear model, for different carrying capacities. The critical funding proportion, $f^*$, in the linear model (solid black curve), and $\phi$ in the nonlinear model ($k = 2$, dashed red curve, $k = 5$, dotted green curve, $k = 15$, dash dotted blue curve, $k = 50$, dashed cyan curve, $k = 10{,}000$ dash dotted magenta curve). The circles a), b) and c) are the values in $(s, f)$ parameter space whose corresponding dynamics are plotted in phase space, with $k = 15$, in Fig. 5. Other parameters are, $m = q = c_O = c_F = \lambda_0 = g = a = 1, b = 2, p_0 = 3.5, \alpha = 3$.

**Case study: Elephant population dynamics if scavenged ivory could fund enforcement**

We consider the case of a government legally trading ivory scavenged from Elephant carcasses. The example is illustrative, based on historical data from the 1990s, and therefore cannot directly inform management today. For example, we will use values for the biological parameters, costs of poaching, and probability of getting caught corresponding mostly to Luangwa Valley, Zambia, in 1992 (Milner-gulland, E.J.; Leader-Williams 1992) and assume that this is the only area supplying ivory to the hypothetical market. The region has been extensively used for an array of theoretical models on elephant poacher dynamics (Lopes 2015; Holden and McDonald-Madden 2017; Holden et al. 2018) and therefore, observing the dynamics from this parameterization helps to understand our models in the context of the greater poacher-population modelling literature.

Wildlife population size, $x(t)$, is the number of elephants at time $t$. We assume an elephant birth rate, 0.33, mortality rate is 0.27 (Conrad and Lopes 2017), and carrying capacity set to 42,984 elephants, corresponding to 2 elephants per km$^2$ in Luangwa Valley reserves (Milner-gulland, E.J.; Leader-Williams 1992). We assume poaching effort, $y(t)$, is the number of gangs poaching at time $t$. We set poacher opportunity cost, $c_O$, to 29.37 USD/gang, based on the daily average sub-Saharan African income of 2.1 USD/person (Lakner and Milanovic 2015), 8 poachers per gang, 7 day expedition time, and 88 USD in expedition expenses (Milner-gulland, E.J.; Leader-Williams 1992). The expected cost of getting caught per anti-poaching patrol per day, was set to 4.69 USD/day/gang/patrol. This was based on the probability of getting caught of 0.05 per expedition, given 13.97 patrols, a penalty of lost income during an average 2-year prison term for each gang member, with discount rate 0.35, and a conviction probability of 0.86 (Leader-Williams et al. 1990; Milner-gulland, E.J.; Leader-Williams 1992; Mastrobuoni and Rivers 2016). The catchability coefficient for elephants was set to 2.56 x 10$^{-5}$ (Milner-gulland, E.J.; Leader-Williams 1992).

The maximum ivory price when there is no ivory on the market and the sensitivity of price to ivory supply are highly uncertain parameters, so we varied these parameters between, 1,000 USD -301,000 USD, and 0-5 respectively. Note that in a study published in 2014, the maximum raw ivory price on the market was 3,740 USD/kg or 28,125 USD/elephant (using 1.88 tusks per elephant). One might typically assume that some of that money goes to the final seller and not the poaching gang, and that the price might be even higher if ivory was rarer on the market. Therefore, it is highly likely that the wide range of 1,000 USD - 301,000 USD per elephant captures the true maximum price the highest bidder would pay for one elephants' worth of ivory. For price sensitivity to supply, $a$, values close to zero represent a shallow drop off from the maximum price, with increasing ivory on the market, and values close to five represent a very steep drop-off. It is estimated that roughly 55 African elephants were poached per day from 2007 to 2014 (Chase et al. 2016), so, for example, $a = 5$ would represent prices of less than a half of one percent of the maximum price, at a 55-elephant/day supply. Therefore, it is likely that the true value of $a$ is somewhere between 0 to 5.

In this example, there is a large region of ($a$, $p_0$) parameter space, where some non-zero proportion of sales revenue funding enforcement (grey area in Fig. 7) is required to save elephants from poaching. In other regions of ($a$, $p_0$) parameter space, funding additional enforcement is not required to save the elephant (white area in Fig. 7). However, the opposite is also possible; there are regions of ($a$, $p_0$) parameter space where legal sales will not save the elephant from poaching, even if all legal sales revenue could fund enforcement (black area in Fig 7). If the government is inefficient at scavenging elephant tusks (e.g. $s = 0.1$, Fig 7ad), or if ivory price is not sensitive to the amount of ivory supplied to the market (small $a$), legal trade is likely ineffective (black area in Fig 7ad). Whereas, legal sales are more likely to achieve positive outcomes for elephants if scavenging is efficient (e.g. $s = 0.5$ or $s = 1$, Fig 7cf) or price is sensitive to ivory supply (large $a$).

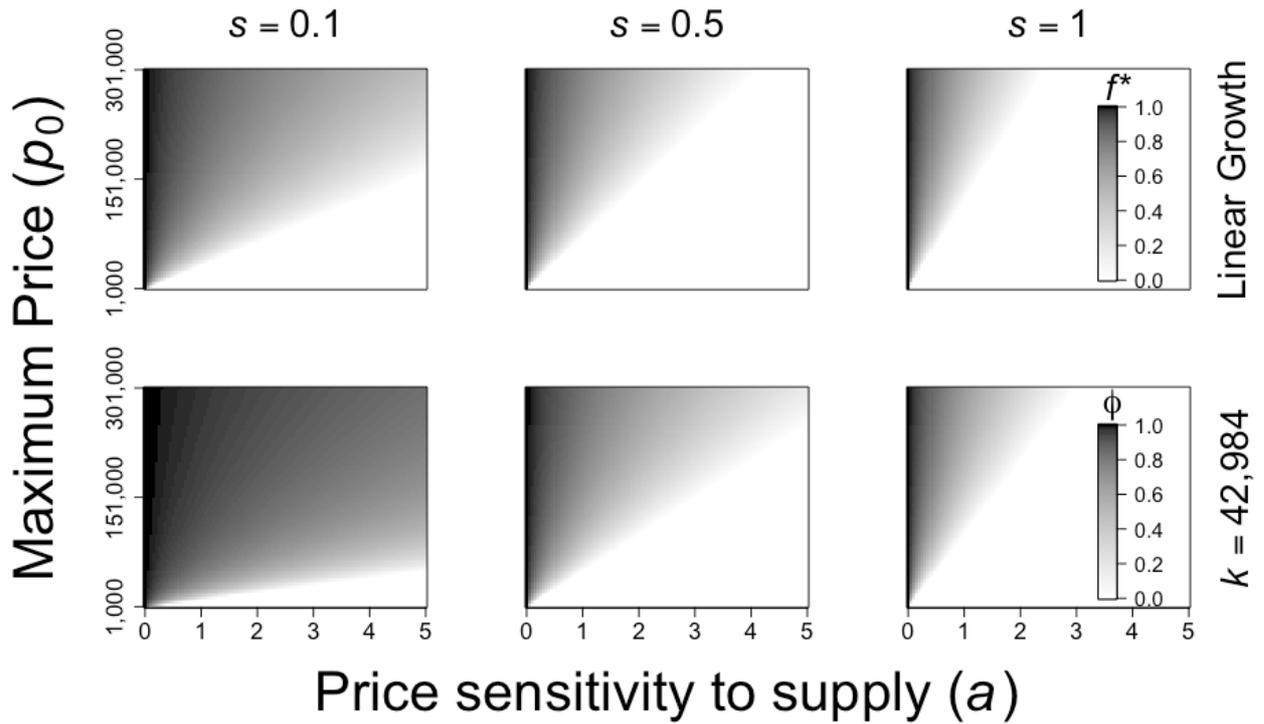

**Fig. 7.** Critical enforcement funding proportion, $f^*$ (in the linear model, row 1), and $\phi$ (in the nonlinear model, row 2) as a function of the price sensitivity, $a$, and maximum price, $p_0$, in the demand curve (equation 3). The black region of the $(a, p_0)$ parameter space is where legal trade is unable to lead to infinite population sizes in the linear model ($f^* > 1$, row 1), or approach carrying capacity in the nonlinear model ($\phi > 1$, row 2). The white region is where legal trade leads to indefinite growth ($f^* < 0$, row 1) or ($\phi < 0$, row 2) carrying capacity regardless as to whether legal trade funds law enforcement or not. The grey region is the critical region where some non-zero proportion of the revenue generated from legal sales must fund enforcement in order for legal trade to lead to indefinitely growing populations or populations at carrying capacity, with darker greys corresponding to higher required funding proportions. The columns correspond to cases where managers are able to scavenge 10% ($s = 0.1$, left column), 50% ($s = 0.5$, middle column), and 100% of ivory ($s = 1$, right column) from elephant carcasses, respectively. Other parameters are $b = 0.33, m = 0.27, q = 2.56 \times 10^{-5}, c_O = 29.37, c_F = 4.69, \lambda_0 = 13.97, g = \frac{1}{150}$, and $k = 42{,}984$ for the nonlinear population growth model.

**Discussion**

In this paper we developed and analyzed a population dynamic model for poached species given a manager scavenging wildlife product from naturally dying organisms to sell and help fund extra anti-poaching law enforcement. The strategy allows the species to benefit from trade, without the manager directly killing any animals. The analysis reveals a critical proportion of legal sales revenue that must go back into funding additional law-enforcement in order for legal trade to alleviate the wildlife population from poaching. Trade funding enforcement is most likely to work if the government can scavenge carcasses efficiently, and if the price of the wildlife product rapidly declines with the amount of product supplied to the market.

Perhaps one of the more prominent results is that equations (11) and (13) show that legal sales funding enforcement will be most beneficial if enforcement efficacy is high. Over the past two years, Uganda and Kenya have been trialling intelligence-based approaches to enforcement (Critchlow et al. 2015; Nguyen et al. 2016) and these countries have seen some rebounds in wildlife populations during this time (Chase et al. 2016). Such intelligence-based strategies would work synergistically with legal sales funding enforcement. However, we must warn that the increased militarization of conservation, and specifically our focus on increasing enforcement funding has considerable drawbacks (Duffy et al. 2019). That said, the models in this paper could be modified to include alternative conservation actions, such as legal sales funding community programs, the creation of protected areas, or ecosystem restoration (Smith et al. 2015).

In our model, scavenging carcasses supplies all legal trade. To simplify the analysis, we only considered the effects of this on poaching and the population size of the poached species. We did not consider effects on humans who live near wildlife (Smith et al. 2015), animal welfare (Derkley et al. 2019), or the ecosystem effects of scavenging carcasses. The latter could be substantial because carcasses are potential disease sources, and food sources for pests, and even threatened species in need of conservation (O'Bryan et al. 2019). Also, for many species, scavenging and selling products from carcasses may not be effective enough to influence poacher incentives or produce meaningful conservation revenue (low *s* in our model). An alternative approach would be to confiscate products from arrested poachers to sell and generate conservation funding. Initial stockpiles of previously confiscated products (Biggs et al. 2016; Braczkowski et al. 2018) could supplement these sales further. While biologically, selling confiscated product is a different management action, including this into our model, and nondimensionalizing the equations, would lead to a structurally similar set of ODEs as presented in equations (7 - 10) and therefore qualitatively similar dynamics.

There are many caveats that must be considered before using our model to directly support or discount the conservation strategy of legal trade funding enforcement. Like the majority of literature in economic theory and wildlife trade, we assumed that a downward sloping demand curve shapes the consumer's willingness to pay for wildlife products (Clark 1990). However, demand curves aggregated across all potential consumers are difficult to measure, and therefore their utility is contentious (Nadal and Aguayo 2014). Alternative price models exist, such as price as a function of wildlife population size, rather than the amount of product on the market (Courchamp et al. 2006; Mason et al. 2012; Holden and McDonald-Madden 2017; Wang and Zhang 2018). This could be more accurate if some dealers accumulate wildlife product as an investment (Mason et al. 2012; Sas-Rolfes et al. 2014), gambling on high product prices in the scenario where the population eventually goes extinct. However, with the exception of a few species, there is more evidence that price is determined by market supply rather than population status (Burgess et al. 2017).

The biggest limitation of the current model is that it ignores corruption. For example, legalizing sales of wildlife products, allows poachers the opportunity to disguise their illegal products as legal (Bennett 2015). Additional complexities also include the possibility of legal trade to reduce stigma associated with purchasing wildlife products, thereby increasing demand from law-abiding consumers (Fischer 2004), and several other uncertainties in how human behaviour may respond to wildlife trade policies ('t Sas-Rolfes et al. 2019). Such models that address more detailed human behaviour may require computer simulation of individual actors (Haas and Ferreira 2016, 2018), operating under probabilistic, game theoretic rules (Missios 2004; Joosten and Meijboom 2018; Glynatsi et al. 2018) and would incorporate poachers competing with the manager by scavenging wildlife product as well.

Despite conservation biologists advocating the possibility for legal sales to fund conservation (Di Minin et al. 2015; Smith et al. 2015), this is the first dynamic model, to our knowledge, to project the consequences of such actions. It therefore represents a new, mathematically well-understood, foundational model for conservation biologists to build off of, by adding the necessary complexities to manage environmental systems. This not only includes poached species, but any species subject to regulated harvest, when naturally deceased organisms have economic value.

**Acknowledgements:**

We thank Duan Biggs who introduced us to the proposed conservation strategy of using legal wildlife trade to increase conservation funding for poached species. Funding was provided to MHH by an ARC DECRA fellowship (DE190101416).